\documentclass[12pt,titlepage]{utarticle}

\usepackage{amsmath}
\usepackage{amsfonts}
\usepackage{amssymb}
\usepackage{amsthm}
\usepackage{mathtools}
\usepackage{graphicx}
\usepackage{color}
\usepackage{ucs}
\usepackage{mathbbol}
\usepackage[utf8x]{inputenc}
\usepackage{hyperref}

\advance\hoffset0.25in

\definecolor{aqua}{rgb}{0, 1.0, 1.0}
\definecolor{fuschia}{rgb}{1.0, 0, 1.0}
\definecolor{gray}{rgb}{0.502, 0.502, 0.502}
\definecolor{lime}{rgb}{0, 1.0, 0}
\definecolor{maroon}{rgb}{0.502, 0, 0}
\definecolor{navy}{rgb}{0, 0, 0.502}
\definecolor{olive}{rgb}{0.502, 0.502, 0}
\definecolor{purple}{rgb}{0.502, 0, 0.502}
\definecolor{silver}{rgb}{0.753, 0.753, 0.753}
\definecolor{teal}{rgb}{0, 0.502, 0.502}

\makeatletter
\newdimen\itex@wd%
\newdimen\itex@dp%
\newdimen\itex@thd%
\def\itexspace#1#2#3{\itex@wd=#3em%
\itex@wd=0.1\itex@wd%
\itex@dp=#2ex%
\itex@dp=0.1\itex@dp%
\itex@thd=#1ex%
\itex@thd=0.1\itex@thd%
\advance\itex@thd\the\itex@dp%
\makebox[\the\itex@wd]{\rule[-\the\itex@dp]{0cm}{\the\itex@thd}}}
\makeatother

\makeatletter
\newif\if@sup
\newtoks\@sups
\def\append@sup#1{\edef\act{\noexpand\@sups={\the\@sups #1}}\act}%
\def\reset@sup{\@supfalse\@sups={}}%
\def\mk@scripts#1#2{\if #2/ \if@sup ^{\the\@sups}\fi \else%
  \ifx #1_ \if@sup ^{\the\@sups}\reset@sup \fi {}_{#2}%
  \else \append@sup#2 \@suptrue \fi%
  \expandafter\mk@scripts\fi}
\def\tensor#1#2{\reset@sup#1\mk@scripts#2_/}
\def\multiscripts#1#2#3{\reset@sup{}\mk@scripts#1_/#2%
  \reset@sup\mk@scripts#3_/}
\makeatother

\makeatletter
\newbox\slashbox \setbox\slashbox=\hbox{$/$}
\def\itex@pslash#1{\setbox\@tempboxa=\hbox{$#1$}
  \@tempdima=0.5\wd\slashbox \advance\@tempdima 0.5\wd\@tempboxa
  \copy\slashbox \kern-\@tempdima \box\@tempboxa}
\def\slash{\protect\itex@pslash}
\makeatother

\def\clap#1{\hbox to 0pt{\hss#1\hss}}

\let\oldroot\root
\def\root#1#2{\oldroot #1 \of{#2}}
\renewcommand{\sqrt}[2][]{\oldroot #1 \of{#2}}

\DeclareSymbolFont{symbolsC}{U}{txsyc}{m}{n}
\SetSymbolFont{symbolsC}{bold}{U}{txsyc}{bx}{n}
\DeclareFontSubstitution{U}{txsyc}{m}{n}

\DeclareSymbolFont{stmry}{U}{stmry}{m}{n}
\SetSymbolFont{stmry}{bold}{U}{stmry}{b}{n}

\DeclareFontFamily{OMX}{MnSymbolE}{}
\DeclareSymbolFont{mnomx}{OMX}{MnSymbolE}{m}{n}
\SetSymbolFont{mnomx}{bold}{OMX}{MnSymbolE}{b}{n}
\DeclareFontShape{OMX}{MnSymbolE}{m}{n}{
    <-6>  MnSymbolE5
   <6-7>  MnSymbolE6
   <7-8>  MnSymbolE7
   <8-9>  MnSymbolE8
   <9-10> MnSymbolE9
  <10-12> MnSymbolE10
  <12->   MnSymbolE12}{}

\makeatletter
\def\re@DeclareMathSymbol#1#2#3#4{%
    \let#1=\undefined
    \DeclareMathSymbol{#1}{#2}{#3}{#4}}
\re@DeclareMathSymbol{\neArrow}{\mathrel}{symbolsC}{116}
\re@DeclareMathSymbol{\neArr}{\mathrel}{symbolsC}{116}
\re@DeclareMathSymbol{\seArrow}{\mathrel}{symbolsC}{117}
\re@DeclareMathSymbol{\seArr}{\mathrel}{symbolsC}{117}
\re@DeclareMathSymbol{\nwArrow}{\mathrel}{symbolsC}{118}
\re@DeclareMathSymbol{\nwArr}{\mathrel}{symbolsC}{118}
\re@DeclareMathSymbol{\swArrow}{\mathrel}{symbolsC}{119}
\re@DeclareMathSymbol{\swArr}{\mathrel}{symbolsC}{119}
\re@DeclareMathSymbol{\nequiv}{\mathrel}{symbolsC}{46}
\re@DeclareMathSymbol{\Perp}{\mathrel}{symbolsC}{121}
\re@DeclareMathSymbol{\Vbar}{\mathrel}{symbolsC}{121}
\re@DeclareMathSymbol{\sslash}{\mathrel}{stmry}{12}
\re@DeclareMathSymbol{\invamp}{\mathrel}{symbolsC}{77}
\re@DeclareMathSymbol{\parr}{\mathrel}{symbolsC}{77}
\makeatother

\makeatletter
\def\Decl@Mn@Delim#1#2#3#4{%
  \if\relax\noexpand#1%
    \let#1\undefined
  \fi
  \DeclareMathDelimiter{#1}{#2}{#3}{#4}{#3}{#4}}
\def\Decl@Mn@Open#1#2#3{\Decl@Mn@Delim{#1}{\mathopen}{#2}{#3}}
\def\Decl@Mn@Close#1#2#3{\Decl@Mn@Delim{#1}{\mathclose}{#2}{#3}}
\Decl@Mn@Open{\llangle}{mnomx}{'164}
\Decl@Mn@Close{\rrangle}{mnomx}{'171}
\makeatother

\makeatletter
\DeclareRobustCommand\widecheck[1]{{\mathpalette\@widecheck{#1}}}
\def\@widecheck#1#2{%
    \setbox\z@\hbox{\m@th$#1#2$}%
    \setbox\tw@\hbox{\m@th$#1%
       \widehat{%
          \vrule\@width\z@\@height\ht\z@
          \vrule\@height\z@\@width\wd\z@}$}%
    \dp\tw@-\ht\z@
    \@tempdima\ht\z@ \advance\@tempdima2\ht\tw@ \divide\@tempdima\thr@@
    \setbox\tw@\hbox{%
       \raise\@tempdima\hbox{\scalebox{1}[-1]{\lower\@tempdima\box
\tw@}}}%
    {\ooalign{\box\tw@ \cr \box\z@}}}
\makeatother

\makeatletter
\def\udots{\mathinner{\mkern2mu\raise\p@\hbox{.}
\mkern2mu\raise4\p@\hbox{.}\mkern1mu
\raise7\p@\vbox{\kern7\p@\hbox{.}}\mkern1mu}}
\makeatother

\newcommand{\gt}{>}
\newcommand{\lt}{<}

\theoremstyle{plain}

\theoremstyle{definition}

\theoremstyle{remark}

\begin{document}

\preprint{
UTTG-21-12\\
TCC-020-12\\
}
\title{On Uncertainties in Successive Measurements}

\author{Jacques Distler and Sonia Paban
     \oneaddress{
      Theory Group and\\
      Texas Cosmology Center\\
      Department of Physics,\\
      University of Texas at Austin,\\
      Austin, TX 78712, USA \\
      {~}\\
      \email{distler@golem.ph.utexas.edu}\\
      \email{paban@physics.utexas.edu}
      }
}

\date{November 20, 2012}

\Abstract{
When you measure an observable, $A$, in Quantum Mechanics, the state of the system changes. This, in turn, affects the quantum-mechanical uncertainty in some non-commuting observable, $B$. The standard Uncertainty Relation puts a lower bound on the uncertainty of $B$ in the initial state. What is relevant for a subsequent measurement of $B$, however, is the uncertainty of $B$ in the \emph{post-measurement} state. We re-examine this problem, both in the case where $A$ has a pure point spectrum and in the case where $A$ has a continuous spectrum. In the latter case, the need to include a finite detector resolution, as part of what it means to measure such an observable, has dramatic implications for the result of successive measurements. Ozawa \cite{Ozawa} proposed an inequality satisfied in the case of successive measurements. Among our results, we show that his inequality is ineffective (can never come close to being saturated). For the cases of interest, we compute a sharper lower bound.
}

\maketitle

\thispagestyle{empty}
\tableofcontents
\vfill
\newpage
\setcounter{page}{1}

\section{Introduction}\label{introduction}

The Uncertainty Principle is one of the signature features of Quantum Mechanics. In popular accounts, it is often described as the principle that measuring some observable ``{}disturbs''{} the system and that this has implications for subsequent observations. As we shall review in \S\ref{purepoint}, the Uncertainty Principle \emph{does not} say anything about \emph{successive} measurements. In fact, formulating a precise statement about the uncertainties in the result of successive measurements (of non-commuting observables) has received, perhaps, less attention than it deserves.

The character of such a statement depends a great deal on whether the observable(s) in question have discrete or continuous spectra. We start with the former case, in \S\ref{purepoint}, and tackle the latter in \S\ref{unbounded}. When the spectrum is continuous, we \emph{need} to include a finite detector resolution as part of what it \emph{means} to measure that observable. This has dramatic (and, to our knowledge, quite novel) implications for the results of successive measurements. In \S\ref{successivexp} we find, for instance, that if one first measures $x$ (with detector resolution, $\sigma_x$) and then measure $p$ (with detector resolution $\sigma_p$), the product of the uncertainties of the measured values is bounded below by
\begin{displaymath}
(\Delta x)^2_{\text{measured}}(\Delta p)^2_{\text{measured}} \geq \frac{1}{4}\left(1+\sqrt{1+4\sigma_x^2\sigma_p^2}\right)^2
\end{displaymath}
which contrasts quite strikingly with the lower bound on the product of quantum-mechanical uncertainties in the initial state: $(\Delta x)^2_{\rho}(\Delta p)^2_{\rho} \geq \frac{1}{4}$. In \S\ref{ArthursKelley}, we extend this formalism to the notion of ``joint measurements" of non-commuting observables.

In  \cite{Ozawa} Ozawa proposed an inequality to be satisfied  in the case of successive measurements. This proposal  remains controversial, as can be seen from the recent papers \cite{Fujikawa:2012ww,DiLorenzo}. Lund and Wiseman \cite{Lund-Wiseman} proposed an experimental test of Ozawa's ideas and Rozema \emph{et al.}~\cite{Rozema:2012sg} recently performed the experiment. As an application, we show in \S\ref{OzawaRel} that Ozawa's inequality is ineffective (can never come close to being saturated). We demonstrate this both for the two-state system relevant to \cite{Lund-Wiseman,Rozema:2012sg} and for operators with continuous spectra ($x$ and $p$). In each case, we compute a sharper lower bound. This illuminates an evident feature of the results of \cite{Rozema:2012sg}: when Rozema \emph{et al.}~measure the quantities appearing on the LHS and RHS of Ozawa's inequality, their results for the LHS nowhere approach the values for the RHS.

Along the way, in \S\ref{Ozawa} we derive an ``intrinsic'' expression for the ``disturbance'' $\eta(B)$, of some observable $B$, which results from measuring some other observable, $A$. In \S\ref{Lund}, we turn to Lund and Wiseman's refinement of Ozawa's proposal, involving a ``weak measurement" of $A$. We show how this can be cast in terms of a \emph{family} of Positive Operator-valued measures (POVMs), and find the corresponding expressions for $\eta(B)$ and $\epsilon(A)$.




The previous version of this manuscript used a less ``efficient" POVM for measurements of $x$ and $p$ (essentially, the POVM for joint measurements, of \S\ref{ArthursKelley}, where one discards the result for one of the jointly-measured observables). The inefficiency of that choice was pointed out to us, in private communication, by A.~Di Lorenzo, whose paper \cite{DiLorenzo} contains results overlapping with those of \S\ref{unbounded}.

\section{Measuring an Observable with a Pure Point Spectrum}\label{purepoint}
\subsection{General setup}\label{setup}

A quantum system is described in terms of a density matrix, $\rho:\, \mathcal{H}\to \mathcal{H}$, which is a self-adjoint, positive-semidefinite trace-class operator, satisfying

\begin{displaymath}
Tr(\rho)=1\qquad 
\end{displaymath}
In the Schr\"o{}dinger picture (which we will use), it evolves unitarily in time

\begin{equation}
\rho(t_2) = U(t_2,t_1) \rho(t_1) {U(t_2,t_1)}^{-1}
\label{unitary}\end{equation}
\emph{except} when a measurement is made.

Consider a self-adjoint operator $A$ (an ``{}observable''{}). We will assume that $A$ has a pure point spectrum, and let $P^{(A)}_i$ be the projection onto the $i^{\text{th}}$ eigenspace of $A$.

When we measure $A$, quantum mechanics computes for us

\begin{enumerate}%
\item A \emph{classical} probability distribution for the values on the readout panel of the measuring apparatus. The moments of this probability distribution are computed by taking traces. The $n^{\text{th}}$ moment is

\begin{displaymath}
{\langle A^n\rangle} = Tr(A^n\rho)
\end{displaymath}
In particular, the variance is

\begin{displaymath}
{(\Delta A)}^2 = Tr(A^2\rho) - {\left(Tr(A\rho)\right)}^2
\end{displaymath}

\item A \emph{change} (which, under the assumptions stated, can be approximated as occurring instantaneously) in the density matrix \cite{vonNeumann},

\begin{equation}
\rho_{\text{after}}\equiv \widehat{\rho}(\rho,A) = \sum_i P^{(A)}_i \rho P^{(A)}_i
\label{rhojump}\end{equation}
Thereafter, the system, described by the new density matrix, $\widehat{\rho}$, again evolves unitarily, according to \eqref{unitary}.

\end{enumerate}
The new density matrix, $\widehat{\rho}$, after the measurement\footnote{Frequently, one wants to ask questions about conditional probabilites: ``{}Given that a measurement of $A$ yields the value $\lambda_1$, what is the probability distribution for a subsequent measurement of \ldots{}''{}. To answer such questions, one typically works with a new (``{}projected''{}) density matrix, $\rho' = \frac{1}{Z} P^{(A)}_1 \rho P^{(A)}_1$, where the normalization factor $Z=Tr(P^{(A)}_1 \rho)$ is required to make $Tr(\rho')=1$. The density matrix $\widehat{\rho}$, however, contains all of the information encoded in $\rho'$. The expectation-value of any observable in the ensemble $\rho'$ may be computed using $\widehat{\rho}$ via
$$
  Tr(\mathcal{O}\rho') = \frac{Tr(P^{(A)}_1 \mathcal{O}P^{(A)}_1 \widehat{\rho})}{Tr(P^{(A)}_1\widehat{\rho})}
$$
The formalism in the main text of this paper is geared to computing joint probability distributions using $\widehat{\rho}$.\label{ftn:conditional}} , can be completely characterized by two properties

\begin{enumerate}%
\item All of the moments of $A$ are the same\footnote{More precisely, the probability measure on $Spec(A)$ is unchanged (see \S\ref{pvm}). This is an important distinction when $A$ is an unbounded operator, as the moments of $A$ generically don'{}t all exist, and so can'{}t be used to characterize the probability distribution.}  as before

\begin{displaymath}
\langle A^n\rangle = Tr(A^n \widehat{\rho})= Tr(A^n \rho)
\end{displaymath}
(In particular, $\Delta A$ is unchanged.) Moreover, for any observable, $C$, which commutes with $A$ ( $[C,A]=0$ ), its moments are also unchanged

\begin{displaymath}
\langle C^n\rangle = Tr(C^n \widehat{\rho})= Tr(C^n \rho)
\end{displaymath}

\item However, the measurement has destroyed all interference between the different eigenspaces of $A$

\begin{displaymath}
Tr([A,B] \widehat{\rho}) = 0, \qquad \forall\, B
\end{displaymath}

\end{enumerate}
Note that it is \emph{really important} that we have assumed a pure point spectrum. If $A$ has a continuous spectrum, then you have to deal with complications both physical and mathematical. Mathematically, you need to deal with the complications of the Spectral Theorem; physically, you have to put in finite detector resolutions, in order to make proper sense of what a ``{}measurement''{} does. Later, we will explain how to deal with those complications.

Now consider two such observables, $A$ and $B$. The Uncertainty Principle gives a lower bound on the product

\begin{equation}
{(\Delta A)}_\rho {(\Delta B)}_\rho \geq \tfrac{1}{2}\left| Tr([A,B]\rho) \right|
\label{uncertainty}\end{equation}
in any state, $\rho$.

The proof, generalizing the usual proof presented for ``{}pure states''{}, is as follows. Let $S= \left(A-{\langle A\rangle}_\rho\mathbb{1}\right)+e^{i\theta}\alpha \left(B-{\langle B\rangle}_\rho\mathbb{1}\right)$ for $\alpha\in\mathbb{R}$. Consider

\begin{displaymath}
Q(\alpha) = Tr(S S^\dagger \rho)
\end{displaymath}
This is a quadratic expression in $\alpha$, which is positive-semidefinite, $Q(\alpha)\geq 0$. Thus, the discriminant must be negative-semidefinite, $D\leq 0$. For $\theta=\pi/2$, this yields the conventional uncertainty relation,

\begin{displaymath}
{(\Delta A)}^2_\rho{(\Delta B)}^2_\rho\geq\frac{1}{4}{\left(Tr(i[A,B]\rho)\right)}^2
\end{displaymath}
For $\theta=0$, it yields

\begin{equation}
{(\Delta A)}^2_\rho{(\Delta B)}^2_\rho\geq{\left(\frac{1}{2}Tr(\{A,B\}\rho)-Tr(A\rho)Tr(B\rho)\right)}^2
\label{altUncertainty}\end{equation}
which is an expression you sometimes see, in the higher-quality textbooks.

As stated, \eqref{uncertainty} is not a statement about the uncertainties in any \emph{actual} sequence of measurements. After all, once you measure $A$, in state $\rho$, the density matrix changes, according to \eqref{rhojump}, to

\begin{equation}
\widehat{\rho}(\rho,A) = \sum_i P^{(A)}_i \rho P^{(A)}_i
\label{rhoprime}\end{equation}
so a subsequent measurement of $B$ is made in a \emph{different state} from the initial one.

If we are interested in the uncertainties associated to successive measurements, the obvious next thing to try is to note that, since the uncertainty of $A$ in the state $\widehat{\rho}(\rho,A)$ is the \emph{same} as in the state $\rho$, and since we are measuring $B$ in the state $\widehat{\rho}$, we can apply the Uncertainty Relation, \eqref{uncertainty} in the state $\widehat{\rho}$, instead of in the state, $\rho$. Unfortunately, $Tr([A,B]\widehat{\rho})=0$, so this leads to an uninteresting lower bound on the product the uncertainties

\begin{equation}
(\Delta A) (\Delta B) = {(\Delta A)}_{\widehat{\rho}} {(\Delta B)}_{\widehat{\rho}} \geq  0
\label{boring}\end{equation}
for a measurement of $A$ immediately followed by a measurement of $B$.

To get a (slightly) more interesting lower bound, let

\begin{equation}
\widehat{B} = \sum_i P^{(A)}_i B P^{(A)}_i
\label{bhat}\end{equation}
and

\begin{displaymath}
M_i = P^{(A)}_i B \left(\mathbb{1} - P^{(A)}_i\right)
\end{displaymath}
We easily compute

\begin{displaymath}
{(\Delta B)}^2_{\widehat{\rho}} = {(\Delta \widehat{B})}^2_{\rho} + \sum_i Tr(M_i M_i^\dagger \rho)
\end{displaymath}
and hence

\begin{equation}
{(\Delta A)}^2_\rho {(\Delta B)}^2_{\widehat{\rho}} = {(\Delta A)}^2_\rho {(\Delta \widehat{B})}^2_{\rho}  + {(\Delta A)}^2_\rho \sum_i Tr(M_i M_i^\dagger \rho)
\label{twoterms}\end{equation}
Since $[\widehat{B},A]=0$, the first term is bounded below (see \eqref{altUncertainty}) by

\begin{equation}
{(\Delta A)}^2_\rho {(\Delta \widehat{B})}^2_{\rho} \geq {\left(\tfrac{1}{2} Tr(\{A,\widehat{B}\}\rho)-Tr(A\rho)Tr(\widehat{B}\rho)\right)}^2
\label{firstterm}\end{equation}
The right hand side is the square of the covariance in the state $\widehat{\rho}$
\begin{equation*}
\mbox{Cov}(A,B)_{\hat{\rho}} \equiv \langle \tfrac{1}{2} \{A,B\} \rangle_{\hat{\rho}} - \langle A\rangle_{\hat{\rho}} \langle B \rangle_{\hat{\rho}}
\end{equation*}
so the inequality \eqref{firstterm} is just the statement that the determinant of the covariance matrix  is non-negative. 

The second term in \eqref{twoterms} is also positive-semidefinite. So, the product of the measured uncertainties, for successive measurements of $A$ then $B$, obeys

\begin{equation}
\begin{split}
{(\Delta A)}^2 {(\Delta B)}^2 &= {(\Delta A)}^2_\rho {(\Delta B)}^2_{\widehat{\rho}}\\
&\geq
{\left(\frac{1}{2} Tr(\{A,\widehat{B}\}\rho)-Tr(A\rho)Tr(\widehat{B}\rho)\right)}^2 + 
{(\Delta A)}^2_\rho \sum_i Tr(M_i M_i^\dagger \rho)
\end{split}
\label{littlebetter}\end{equation}
It is conventional to define a quantity

\begin{equation}
\eta(B)^2 = Tr\Bigl[\bigl({(B-\widehat{B})}^2+\widehat{B^2}-\widehat{B}^2\bigr)\rho\Bigr]
\label{disturbance}\end{equation}
called ``{}the \emph{disturbance} in $B$''{} (due to the measurement of $A$). Here $\widehat{B}$ is defined as in \eqref{bhat} (and similarly for $\widehat{B^2}$), and the trace is taken in the initial (pre-measurement) state, $\rho$. The combination $\widehat{B^2}-\widehat{B}^2$, which appears in \eqref{disturbance}, is exactly the positive semi-definite operator, $\sum_i M_i M_i^\dagger$, which appeared in \eqref{littlebetter}.

\subsection{The two-state system}\label{twostate}

For the classic case of a 2-state system, with $A=J_x$ and $B=J_y$, we see that $\widehat{B}\equiv 0$, and the product of uncertainties is entirely given by the second term of \eqref{twoterms}.

The most general density matrix for the 2-state system is parametrized by the unit 3-ball\footnote{More generally, for an $n$-dimensional Hilbert space, $\mathcal{H}$, the set of density matrices is an $(n^2-1)$-dimensional convex set $S\subset \mathbb{R}^{n^2-1}$. The $(n^2-2)$-dimensional boundary of $S$ consists of density matrices on proper subspaces $\mathcal{H}'\subset \mathcal{H}$. The set of pure states is a closed $(2n-2)$-dimensional subset of the boundary consisting of density matrices of rank-1. This $\mathbb{C P}^{n-1}$ consists of the \emph{extreme points} of $S$ \cite{Freed}.} 

\begin{displaymath}
\rho = \frac{1}{2}\left(\mathbb{1}+ \vec{a}\cdot\vec{\sigma}\right), \qquad  \vec{a}\cdot\vec{a}\leq 1
\end{displaymath}
The points on the boundary $S^2=\{\vec{a}\cdot\vec{a}=1\}$ correspond to pure states.

Upon measuring $A=J_x\equiv\tfrac{1}{2}\sigma_x$, the density matrix after the measurement is

\begin{displaymath}
\widehat{\rho} = \frac{1}{2}\left(\mathbb{1}+ a_x\sigma_x\right)
\end{displaymath}
and, for a subsequent measurement of $J_y$,

\begin{displaymath}
{(\Delta J_x)}^2_\rho{(\Delta J_y)}^2_{\widehat{\rho}} = \frac{1}{16}(1-a_x^2)
\end{displaymath}
as ``{}predicted''{} by \eqref{twoterms}.

More generally, consider $B = \vec{b}\cdot \vec{J}$. The quantum-mechanical uncertainty in the initial state is given by

\begin{displaymath}
{(\Delta B)}^2_\rho = \frac{1}{4}\left[b^2-{(\vec{a}\cdot\vec{b})}^2\right]
\end{displaymath}
After measuring $A=J_x$, the quantum-mechanical uncertainty in the state $\widehat{\rho}$ is given by

\begin{displaymath}
{(\Delta B)}^2_{\widehat{\rho}} = \frac{1}{4}\left[b^2-a_x^2b_x^2\right]
\end{displaymath}
so the change in the uncertainty,

\begin{equation}
{(\Delta B)}^2_{\widehat{\rho}} - {(\Delta B)}^2_\rho = \frac{1}{4}\left[{(\vec{a}\cdot\vec{b})}^2 -a_x^2b_x^2 \right]
\label{disturbing}\end{equation}
can be arbitrarily negative. If we fix $b^2=1$, and take $\vec{a}=\frac{1}{\sqrt{2}}(1,1,0)$, corresponding to the pure state

\begin{displaymath}
\rho=\frac{1}{2}\begin{pmatrix}1&e^{-i\pi/4}\\ e^{i\pi/4}&1\end{pmatrix}
\end{displaymath}
then the minimum for \eqref{disturbing} is achieved for $\vec{b}=\frac{1}{\sqrt{2}} (1,-1,0)$ (i.e., $B=\frac{1}{\sqrt{2}}(J_x-J_y)$):

\begin{equation}
{(\Delta B)}^2_{\widehat{\rho}} - {(\Delta B)}^2_\rho = -\frac{1}{16}
\label{coffin}\end{equation}
Nevertheless, the disturbance,

\begin{displaymath}
{\eta(B)}^2 = \frac{1}{2}(b_y^2+b_z^2) = \frac{1}{4} \gt 0
\end{displaymath}
Note however that, for the 2-state system, $\eta(B)$ depends only on the relative orientation of $A=J_x$ and $B=\vec{b}\cdot\vec{J}$ and \emph{not at all} on the state, $\rho$, in which it is measured. We will return to this point in \S\ref{Ozawa}.

Finally, note that what is relevant to the successive measurements is the covariance in the state $\widehat{\rho}$, rather than in the state $\rho$. These are different:

\begin{equation}
\begin{split}
\text{Cov}(A,B)_{\rho} =& \frac{1}{4} \left( b_x - a_x \, \vec{a} \cdot \vec{b} \right)\\
\text{Cov}(A,B)_{\hat {\rho}} =& \frac{1}{4}( 1- a_x^2) \, b_x
\end{split}
\end{equation}

Before we turn to the measurement of observables with continuous spectra, we need to make a disgression on the Spectral Theorem, Projection-Values Measures, and the more general concept of Positive Operator-Valued Measures.

\section{Projection-valued and Positive operator-valued measures}\label{pvmpovm}

\subsection{PVMs and observables}\label{pvm}

For an observable, $A$, with spectrum $X=Spec(A)$, it is rather \emph{awkward} (and, in the unbounded case, generically \emph{impossible}) to describe a probability distribution \emph{implicitly}, in terms of its moments, $\langle A^n\rangle$. Instead, we would like to directly give a \emph{formula} for the desired probability measure on $X$.

That is what the formalism of projection-valued measures (PVMs) affords us (see, e.g., \cite{Mackey}). A projection-valued measure on $X$ is a rule which assigns to every Borel subset $S\subset X$ a projection operator $\pi_S$ on $\mathcal{H}$ satisfying the obvious properties

\begin{itemize}%
\item $\pi_X =\mathbb{1}$.
\item $\pi_\emptyset = 0$.
\item $\pi_{S\cap S'} = \pi_S \pi_{S'}$.
\item $\pi_{S_1\cup S_2\cup\dots}=\pi_{S_1} + \pi_{S_2}+\dots$ whenever $S_i\cap S_j =\emptyset$.

\end{itemize}
The Spectral Theorem is the statement that there is a 1-1 correspondence between self-adjoint operators, $A$, with $Spec(A)=X$ and projection-valued measures on $X$.

For $\mathcal{H}=\mathcal{L}^2(\mathbb{R})$ and $A=\mathbf{x}$, the corresponding projection-valued measure is

\begin{equation}\label{xPVM}
(\pi_S \psi) (x) =\begin{cases}\psi(x)&x\in S\\0&\text{otherwise}\end{cases}
\end{equation}
For $A=\mathbf{p}$, the corresponding projection-valued measure (slightly schematically) is

\begin{displaymath}
(\pi_S \psi) (x) =\int d y \int_S \frac{d k}{2\pi} e^{i k (x-y)}\psi(y)
\end{displaymath}
With the Spectral Theorem in hand, it is easy to give a more explicit description of the probability measure on $X=Spec(A)$. It is the measure which assigns to each Borel subset, $S\subset Spec(A)$, the probability

\begin{equation}\label{pvmprob}
P(S) = Tr(\pi_S \rho)
\end{equation}

Unfortunately, when $X$ is a continuous spectrum, there's no generalization of von Neumann's formula \eqref{rhojump} for the change in the state, due to the measurement. Indeed, no formula is possible until we introduce the notion of a finite detector resolution. But, in order to do that, we need a more expansive notion of what is ``measurable" in Quantum Mechanics, which brings us to the concept of a Positive Operator-Valued Measure.

\subsection{POVMs and ``{}noisy''{} measurements}\label{povm}

The Spectral Theorem establishes a correspondence between self-adjoint operators and projection-valued measures. Given a state, $\rho$, the projection-valued measure gives us a probability measure on $X=Spec(A)$. But a projection-valued measure is \emph{not} the most \emph{general} way to manufacture such a probability measure on $X$, from a state, $\rho$.  

In discusssing observables with a pure point spectrum, we have assumed that our measuring apparati are ``{}ideal.'' That is, we have assumed that the measured value of $A$ was an accurate reflection of the quantum state, $\rho$, of the system. One extension of this analysis would include a discussion of imperfect measuring apparati. For instance, even with a discrete spectrum, our measuring apparatus might be unable to distinguish between two closely-spaced eigenvalues $\lambda_i,\lambda_j$ (this is a particularly pressing concern when the spectrum of eigenvalues has an accumulation point).

In this case, the measurement need not destroy the interference between these eigenspaces. To account for this, we replace the pair of projections $P^{(A)}_i,P^{(A)}_j$ with a single projection $P^{(A)}_{i j} = P^{(A)}_i + P^{(A)}_j$ in von Neumann'{}s formula \eqref{rhojump} for the change in the state as a result of the measurement.

More insidiously, the measuring apparatus may not be perfectly classical in its behaviour. The different values for the pointer variable(s) may not decohere. In that case, the ``{}measurement''{} produces an entangled state of the measuring apparatus and the system under study. For some purposes, this is actually an \emph{advantage} and much current effort is devoted to exploiting the possibilities this affords.

To account for these and other, more general, effects, one must leave the world where observables are self-adjoint operators (equivalently, projection-valued measures) and entertain a wider class of observables (see, e.g., \cite{peres} for an introduction), corresponding to positive operator-valued measures (POVMs).

Neumark'{}s Theorem guarantees that a POVM can always be represented as a PVM on some larger Hilbert space. When the resulting PVM has a pure point spectrum, the POVM is a countable collection of self-adjoint positive-semidefinite operators, $\{F_i\}$, satisfying

\begin{displaymath}
\sum_i F_i = \mathbb{1}
\end{displaymath}
(Here $Spec(F_i)\subset [0,1]$, whereas projection operators had $Spec(\pi_i)=\{0,1\}$.) The probability of measuring the value $\lambda_i$ is given, in precise analogy with \eqref{pvmprob}, by
$$
  P(\lambda_i) = Tr(F_i \rho)
$$

Unfortunately, even in this simplest case, the replacement for \eqref{rhojump} is not determined by the POVM alone. We can write

\begin{equation}
F_i = L^\dagger_i L_i
\label{povmdecomp}\end{equation}
and \eqref{rhojump} is replaced by

\begin{equation}
\rho_{\text{after}} \equiv \widehat{\rho} = \sum_i L_i \rho L^\dagger_i
\label{povmcollapse}\end{equation}
But there are an infinite number of solutions to \eqref{povmdecomp}, and hence an ambiguity in \eqref{povmcollapse}. Most authors take the extra data of a choice of $L_i$, satisfying \eqref{povmdecomp}, as part of the \emph{prescription} of the POVM, and we will follow that tradition\footnote{There is another, not always appreciated, difference between the PVM and POVM case. In footnote \ref{ftn:conditional}, we saw that, in the PVM case, all the information required to compute conditional probabilities was encoded in the post-measurement state, $\widehat{\rho}$. That is no longer true in the POVM case.}.

We will find that measuring --- with finite detector resolution --- an operator with a continuous spectrum, can be cast as a particular class of POVMs.

\section{Measuring Unbounded Operators with Continuous Spectra}\label{unbounded}
\subsection{Successive measurements of $x$ and $p$}\label{successivexp}

Let'{}s go straight to the worst-case, of an unbounded operator, with $Spec(A)=\mathbb{R}$. Such an operator has no eigenvectors at all. What happens when we measure such an observable? Clearly, the two conditions which characterized the change in the density matrix, in the case of a pure point spectrum,

\begin{enumerate}%
\item The probability distribution for measured values of $A$ has moments given by\\ $Tr(A^n \widehat{\rho})= Tr(A^n \rho)$.
\item $Tr([A,B] \widehat{\rho}) = 0, \qquad \forall\, B$

\end{enumerate}
are going to have to be modified. The second condition clearly can'{}t hold for all choice of $B$, in the continuous-spectrum case (think $A=\mathbf{x}$ and $B=\mathbf{p}$). As to the first condition, the probability distribution, for the \emph{measured} values of $A$, \emph{depend} on the detector resolution. But the state $\rho$ knows nothing about that. Instead, we will argue that the probability distribution for the measured values of $A$ is given by a particular POVM.

To keep things simple, let'{}s specialize to $\mathcal{H}=\mathcal{L}^2(\mathbb{R})$ and $A=\mathbf{x}$. Imagine a detector which can measure the particle'{}s position with a resolution, $\sigma_x$. We expect this detector resolution to add, in quadrature, to the quantum-mechanical uncertainty, of $\mathbf{x}$ in the state $\rho$, to give the uncertainty of the \emph{measured} value of $x$.

A POVM which implements this is the following. Define the positive semi-definite operator
\begin{equation}
F_{x_0} = L_{x_0}^\dagger L_{x_0}
\end{equation}
where $L_{x_0}$ is the operator
\begin{equation}\label{Lxdef}
L_{x_0}:\quad \psi(x)\mapsto f(x-x_0) \psi(x)
\end{equation}
If the function $f$ satisfies the normalization condition
\begin{equation}\label{fnorm}
1= \int d u\, |f(u)|^2
\end{equation}
then the $F_{x_0}$ obey the completeness relation
\begin{equation}
\int d x_0\, F_{x_0} = \mathbb{1}
\end{equation}
and define a POVM which assigns to a Borel subset $S\subset Spec(\mathbf{x})$ the positive operator
\begin{equation}\label{xPOVM}
S\mapsto \mathcal{F}_S = \int_S dx_0 F_{x_0}
\end{equation}
Because it simplifies several of the formul\ae\ which follow, we will always assume that this ``acceptance function" is an even function: $f(u)=f(-u)$.

When we measure $x$, with this finite-resolution detector, the density matrix changes, according to the generalization of von Neumann's formula, \eqref{povmcollapse}:
\begin{equation}\label{contpovmcollapse}
\widehat{\rho} = \int d x_0\, L_{x_0}\, \rho\, L_{x_0}^\dagger
\end{equation}
If $\rho$ is represented by the integral kernel, $K(x,y)$, then the state after the measurement, $\widehat{\rho}$, is represented by the kernel
\begin{equation}\label{newK}
\widehat{K}(x,y) = \int d x_0\, f(x-x_0)\overline{f}(y-x_0)\, K(x,y)
\end{equation}
If the acceptance function, $f(u)$, is sharply-peaked near $u=0$, then \eqref{newK} has the desired effect of suppressing the off-diagonal elements of $\widehat{K}(x,y)$.

The quantum-mechanical probability distribution for $x$ is unaltered by \eqref{contpovmcollapse}. Indeed, all the moments
\begin{displaymath}
Tr(\mathbf{x}^k\widehat{\rho})= Tr(\mathbf{x}^k\rho)
\end{displaymath}
and, in particular, $(\Delta x)^2_{\widehat{\rho}}=(\Delta x)^2_{\rho}$. However, the probability distribution for the \emph{measured} values of $x$ is given, not by the PVM \eqref{xPVM}, but by the POVM \eqref{xPOVM}, which \emph{is} sensitive to the detector resolution
\begin{equation}\label{quadrature}
\begin{split}
(\Delta x)^2_{\text{measured}} &\equiv \int dx_0\, x_0^2\, Tr(F_{x_0}\rho) - \left(\int d x_0\, x_0\, Tr(F_{x_0}\rho)\right)^2\\
&= (\Delta x)^2_{\rho} + \sigma_x^2
\end{split}
\end{equation}
where
\begin{equation}
\sigma_x^2 = \int d u\, u^2 |f(u)|^2
\end{equation}
Thus we have the two properties we sought:
\begin{itemize}
\item \eqref{newK} expresses our expectation that measuring $x$ suppresses the off-diagonal elements of the density matrix, $\widehat{\rho}$, after the measurement: $\widehat{K}(x,y)\to 0$ for $|x-y|\gg \sigma_x$.
\item \eqref{quadrature} expresses our expectation that the detector resolution, $\sigma_x$ should add in quadrature with the quantum-mechanical uncertainty to give the \emph{measured} uncertainty.
\end{itemize}

While quantum-mechanical probability distribution for $x$ was unaffected by the measurement, the probability distribution for $p$ is altered by \eqref{contpovmcollapse}. The first few moments are
\begin{equation}
\begin{gathered}
Tr(\mathbf{p}\widehat{\rho})=Tr(\mathbf{p}\rho)\\
Tr(\mathbf{p}^2\widehat{\rho})=Tr(\mathbf{p}^2\rho)+\eta(p)^2
\end{gathered}
\end{equation}
where
\begin{displaymath}
\eta(p)^2 = -\int du \overline{f}(u) f''(u)= \int du\, |f'(u)|^2
\end{displaymath}
In particular, the quantum-mechanical uncertainty increases as a result of the measurement
\begin{equation}\label{changedundertainty}
(\Delta p)^2_{\widehat{\rho}}=(\Delta p)^2_{\rho}+\eta(p)^2
\end{equation}

An example, which will prove to be most useful, is a Gaussian acceptance function
\begin{equation}\label{GaussianAcceptance}
f(u) = \frac{1}{(\sigma_x\,\sqrt{2\pi})^{1/2}}e^{-u^2/4\sigma_x^2}
\end{equation}
For \eqref{newK}, we obtain
\begin{equation}
\widehat{K}(x,y) = e^{-(x-y)^2/8\sigma_x^2} K(x,y)
\end{equation}
and
\begin{equation}\label{best}
\eta(\mathbf{p})^2 = \frac{1}{4\sigma_x^2}
\end{equation}
The advantage of the Gaussian acceptance function is that, for given detector resolution, $\sigma_x$, it minimizes the change \eqref{changedundertainty} in the quantum-mechanical uncertainty in $p$.

Of course, the quantum-mechanical uncertainty in $p$ is not quite what we measure. We need to impose a finite detector resolution for $p$, as well. Again, this corresponds to a POVM. Let us take a ``Gaussian" detector\footnote{This is just the momentum space version of the Gaussian detector we described for measuring $x$.}
\begin{equation}\label{pGausian}
  \check{L}_{k_0}:\quad \psi(x)\mapsto \left(\frac{\sigma_p}{\sqrt{\pi/2}}\right)^{1/2}\int dy\, e^{-\sigma_p^2(x-y)^2+i k_0(x-y)}\, \psi(y)
\end{equation}
and
\begin{equation}
\check{F}_{k_0} = \check{L}_{k_0}^\dagger \check{L}_{k_0} :\quad \psi(x)\mapsto \int dy\, e^{-\tfrac{\sigma_p^2}{2}(x-y)^2+i k_0(x-y)}\, \psi(y)
\end{equation}
which obey the completeness relation
\begin{equation}
\int\frac{dk_0}{2\pi}\, \check{F}_{k_0}=\mathbb{1}
\end{equation}
The uncertainty in the measured value
\begin{equation}
\begin{split}
(\Delta p)^2_{\text{measured}}&=
\int \frac{dk_0}{2\pi}\, k_0^2\, Tr(\check{F}_{k_0} \widehat{\rho}) - \left(\int \frac{dk_0}{2\pi}\, k_0\, Tr(\check{F}_{k_0} \widehat{\rho})\right)^2\\
&= (\Delta p)^2_{\widehat{\rho}} + \sigma_p^2
\end{split}
\end{equation}
Using \eqref{changedundertainty},
\begin{equation}\label{Deltapmeasured}
\begin{split}
(\Delta p)^2_{\text{measured}}&= (\Delta p)^2_{\widehat{\rho}}+\sigma_p^2\\
&=(\Delta p)^2_{\rho} +\eta(p)^2 + \sigma_p^2\\
&\geq (\Delta p)^2_{\rho} + \frac{1}{4\sigma_x^2} +  \sigma_p^2 
\end{split}
\end{equation}
where, in the last line, we used that $\eta(p)^2$ is minimized for the Gaussian detector, \eqref{best}.

From this, we see the product of uncertainties --- for successive measurements, first of $x$ then of $p$ --- obeys
\begin{equation}\label{xplowerbound}
\begin{split}
(\Delta x)^2_{\text{measured}}(\Delta p)^2_{\text{measured}}&=
\left((\Delta x)^2_{\rho}+\sigma_x^2\right)\left((\Delta p)^2_{\rho}+\eta(p)^2+\sigma_p^2\right)\\
&\geq \frac{1}{2}+ \left(\frac{1}{4\sigma_x^2}+\sigma_p^2\right)(\Delta x)^2_{\rho}
+ \sigma_x^2 (\Delta p)^2_{\rho} +\sigma_x^2\sigma_p^2\\
&\geq \frac{1}{4}\left(1+\sqrt{1+4\sigma_x^2\sigma_p^2}\right)^2
\end{split}
\end{equation}
where we used that $(\Delta x)^2_{\rho} (\Delta p)^2_{\rho}\geq\tfrac{1}{4}$. The last inequality is saturated by choosing $\rho$ to be a pure state, consisting of a Gaussian wave packet, with very carefully chosen width:
\begin{displaymath}
(\Delta x)^2_\rho= \frac{\sigma_x^2}{\sqrt{1+4\sigma_x^2\sigma_p^2}},\qquad (\Delta p)^2_\rho= \frac{\sqrt{1+4\sigma_x^2\sigma_p^2}}{4\sigma_x^2}
\end{displaymath}

One can consider other acceptance functions in \eqref{Lxdef}. One nice choice is

\begin{displaymath}
f(x) = \frac{s}{1+\frac{\cosh(\alpha x)}{\cosh(\alpha b)}},\qquad s^2 =\frac{\alpha}{2\coth^2(\alpha b)(\alpha b \coth(\alpha b) -1)}
\end{displaymath}
which asymptotes to a square pulse (supported on $x\in[-b,b]$ ) in the limit $\alpha\to \infty$. Doing the requisite integrals is a little harder, in this case, but still eminently doable. We easily compute
\begin{displaymath}
\widehat{\mathcal{P}^n}(x,y) = G(x-y) {\left(-i\frac{\partial}{\partial x}\right)}^n\delta(x-y)
\end{displaymath}
where $\widehat{\mathcal{P}}$ is the integral kernel for the operator $\hat{p}$ and
\begin{displaymath}
\begin{split}
G(u) &= \int d w f(w) f(w-u)\\
&= \frac{\alpha\sinh(\alpha b)}{2(\alpha b \coth(\alpha b) -1)} \frac{(b-u/2)\sinh\alpha(b+u/2)-(b+u/2)\sinh\alpha(b-u/2)}{\sinh(\alpha u/2) \sinh\alpha(b+u/2)\sinh\alpha(b-u/2)}\\
&= 1 -\frac{\alpha^2}{12}\left(\frac{1}{\alpha b \coth(\alpha b) -1}-\frac{3}{\sinh^2(\alpha b)}\right) u^2 + O(u^4)
\end{split}
\end{displaymath}
and hence
\begin{displaymath}
\widehat{\mathbf{p}}=\mathbf{p},\qquad \widehat{\mathbf{p}^2}=\mathbf{p}^2 +\eta(p)^2 \mathbb{1},\qquad \eta(p)^2 = \frac{\alpha^2}{6}\left(\frac{1}{\alpha b \coth(\alpha b) -1}-\frac{3}{\sinh^2(\alpha b)}\right)
\end{displaymath}
Similarly,
\begin{displaymath}
\widehat{\mathbf{x}}=\mathbf{x},\qquad \widehat{\mathbf{x}^2}=\mathbf{x}^2+\sigma_x^2\mathbb{1}
\end{displaymath}
where
\begin{displaymath}
\sigma_x^2 =\int du\, u^2 f(u)^2 = b^2 \left(1-\frac{2\alpha b \cosh(\alpha b)}{3(\alpha b \cosh(\alpha b) -\sinh(\alpha b))} +\frac{\pi^2}{3(\alpha b)^2}\right)
\end{displaymath}
These yield the shifts in the uncertainties due to measuring $x$, with this smoothed-square-wave detector:
\begin{displaymath}
(\Delta p)^2_{\widehat{\rho}}=(\Delta p)^2_{\rho}+\eta(p)^2,\qquad (\Delta x)^2_{\mbox{measured}}=(\Delta x)^2_{\rho}+\sigma_x^2
\end{displaymath}
As you can see, for any value of the dimensionless parameter, $\alpha b$, the smoothed-square-wave detector satisfies $\eta(p)^2 > 1/4\sigma_x^2$.

\subsection{Arthurs-Kelly joint measurements}\label{ArthursKelley}

The notion of a POVM generalizes to measure spaces, $X$,  more complicated than $X\subset\mathbb{R}$. The most obvious generalization is to joint measurements, as discussed by Arthurs and Kelly \cite{Arthurs-Kelly}. Here, $X=Spec(\mathbf{x})\times Spec(\mathbf{p})=\mathbb{R}^2$. Define
\begin{equation}\label{Lx0k0def}
 L_{x_0,k_0}: \psi(x) \mapsto \int dy\, h(x-x_0,y-x_0)\, e^{i k_0(x-y)}\, \psi(y)
\end{equation}
where $h(u,v)$ is normalized 
\begin{displaymath}
1 = \int du dv\, |h(u,v)|^2
\end{displaymath}
so that $F_{x_0,k_0} =  L_{x_0,k_0}^\dagger  L_{x_0,k_0}$ obeys the completeness relation
\begin{displaymath}
 \int \frac{dx_0 dk_0}{2\pi}\, F_{x_0,k_0} = \mathbb{1}
\end{displaymath}
and therefore defines a POVM on $X=Spec(\mathbf{x})\times Spec(\mathbf{p})$ via
\begin{equation}\label{AKpovm}
 S\mapsto \mathcal{F}_S = \int_S \frac{dx_0 dk_0}{2\pi}\, F_{x_0,k_0}
\end{equation}
Since we'll have recourse to them, let us denote the moments of this probability distribution by
\begin{equation}\label{AKmoments}
\begin{split}
\llangle x^k p^l\rrangle &= \int \frac{dx_0 dk_0}{2\pi}x_0^k k_0^l\, Tr(F_{x_0,k_0}\rho)\\
&=\int du dv dx\, \overline{h}(u,v)\, (x-v)^k\left.\left(-i\frac{\partial}{\partial s}\right)^l\right\vert_{s=0} \bigl(h(u,v+s)K(x+s,x)\bigr)
\end{split}
\end{equation}

Arthurs and Kelly's choice corresponds to
\begin{equation}\label{AKfactorized}
h(u,v)=g(u) \overline{f}(v)
\end{equation}

where $f(v)$ is a Gaussian
\begin{equation}\label{AKgaussian}
f(v) = \frac{1}{(b\,\,\sqrt{\pi})^{1/2}}\, e^{-v^2/2b^2}
\end{equation}
and $g(u)$ is an (a-priori independent) acceptance function, normalized as in \eqref{fnorm}. The resulting $F_{x_0,k_0}$,
\begin{equation}\label{AK-F}
F_{x_0,k_0}:\quad \psi(x)\mapsto \int dy f(x-x_0)\overline{f}(y-x_0) e^{ik_0(x-y)} \psi(y)
\end{equation}
is independent of the choice $g(u)$. However, the density matrix, after the joint measurement,
\begin{displaymath}
\widehat{\rho}= \int \frac{dx_0 dk_0}{2\pi}\, L_{x_0,k_0}\,\rho\, L_{x_0,k_0}^\dagger
\end{displaymath}
 \emph{does} depend on that choice. If $K(x,y)$ is the integral kernel representing $\rho$, then the integral kernel representing $\widehat{\rho}$ is
\begin{equation}\label{jointfinal}
\begin{split}
\widehat{K}(x,y) &= \int du dv\, h(x-v,x+u-v)\overline{h}(y-v,y+u-v)K(x+u,y+u) \\
&=\int du dv\, g(x-v)\overline{g}(y-v)\overline{f}(x+u-v)f(y+u-v)\, K(x+u,y+u)
\end{split}
\end{equation}
A peculiar feature of \eqref{AKfactorized} is that the $F_{x_0,k_0}$, in \eqref{AK-F}, are projection operators\footnote{But note that the $F_{x_0,k_0}$ are not orthogonal: $F_{x_0,k_0}F_{x'_0,k'_0}\neq 0$, so \eqref{AKpovm} is still a POVM, not a PVM.}. Moreover, when $g=f$, the $L_{x_0,k_0}$ are also projection operators. 

The measured uncertainties in $x$ and $p$, for this joint measurement, are
\begin{equation}
\begin{aligned}
(\Delta x)^2_{\text{measured}}&=\llangle x^2\rrangle-\llangle x\rrangle^2&\qquad\quad
(\Delta p)^2_{\text{measured}}&=\llangle p^2\rrangle-\llangle p\rrangle^2\\
&= (\Delta x)^2_\rho + \sigma_x^2& &= (\Delta p)^2_\rho + \sigma_p^2
\end{aligned}
\end{equation}
where
\begin{equation}\label{jointdisturbances}
\begin{split}
\sigma_x^2 &= \int du\, u^2\, |f(u)|^2\\
\sigma_p^2 &= - \int du\, \overline{f}(u) f''(u) = \int du\, |f'(u)|^2
\end{split}
\end{equation}
and, as usual, we've assumed that $f(u)=f(-u)$. The Arthurs-Kelly choice \eqref{AKgaussian} of a Gaussian, yields $\sigma_x^2=b^2/2$ and $\sigma_p^2= 1/2b^2$ and hence
\begin{equation}\label{measjoint}
(\Delta x)^2_{\text{measured}} = (\Delta x)^2_{\rho}+ \frac{b^2}{2},\qquad
(\Delta p)^2_{\text{measured}} = (\Delta p)^2_{\rho}+ \frac{1}{2 b^2}
\end{equation}
So, for the joint measurement, we have
\begin{equation}
\begin{split}
(\Delta x)^2_{\text{measured}}(\Delta p)^2_{\text{measured}}
&\geq \frac{1}{2} + (\Delta x)^2_{\rho}\, \frac{1}{ 2 \, b^2} + (\Delta p)^2_{\rho}\,\frac{b^2}{2}\\
&\geq 1
\end{split}
\end{equation}
where the last inequality is saturated for a pure state, $\rho$, which is a Gaussian wave packet of very carefully chosen width:
\begin{displaymath}
(\Delta x)^2_{\rho} = \frac{b^2}{2},\quad (\Delta p)^2_{\rho} = \frac{1}{2b^2}
\end{displaymath}
Comparing with \eqref{xplowerbound}, we see that this is akin to the limit $\sigma_p\to 0$ for the successive measurements (so that the measured uncertainties \eqref{quadrature} and \eqref{Deltapmeasured} approach those of \eqref{measjoint}).

It is illuminating to compare the state after the successive measurements with that after the joint measurement. If we first measure $x$ with the Gaussian detector \eqref{GaussianAcceptance} and then measure $p$ with the Gaussian detector \eqref{pGausian}, the final state, $\rho_{\text{final}}$, is given by the integral kernel
\begin{equation}\label{successivefinal}
K_{\text{final}}(x,y) = e^{-(x-y)^2/8\sigma_x^2}\frac{\sigma_p}{\sqrt{\pi/2}}\int du\, e^{-2\sigma_p^2 u^2}\, K(x+u,y+u)
\end{equation}
which suppresses the off-diagonal elements, as before, and smears the near-diagonal elements against a Gaussian. If you wish, we can think of this successive measurement as a POVM with $L_{x_0,k_0} =\check{L}_{k_0} L_{x_0}$
\begin{equation}
L_{x_0,k_0} =\check{L}_{k_0} L_{x_0}: \psi(x) \mapsto \left(\frac{\sigma_p}{\sigma_x \pi} \right)^{1/2} \int \, dy \, e^{-(y-x_0)^2/{ 4 {\sigma_x}^2} + \sigma_p^2 (x-y)^2} \, e^{i k_0 (x-y)} \psi(y) 
\end{equation}
where, in contrast to Arthurs and Kelly \eqref{AKfactorized}, $h(u,v)= \left(\tfrac{\sigma_p}{\pi\sigma_x}\right)^{1/2} \exp-[v^2/4\sigma_x^2+\sigma_p^2 (u-v)^2]$ doesn't factorize.

The resulting final state, \eqref{successivefinal} yields the net disturbance (as a result of both measurements)
\begin{equation}
\eta(p)^2 = \frac{1}{4\sigma_x^2},\qquad \eta(x)^2 = \frac{1}{4\sigma_p^2}
\end{equation}
For the joint measurement, the final state \eqref{jointfinal} depends on the choice of $g(u)$, which we have heretofore left arbitrary. If we choose a Gaussian
\begin{displaymath}
g(u) = \frac{1}{(a\,\,\sqrt{\pi})^{1/2}}\, e^{-u^2/2a^2}
\end{displaymath}
then \eqref{jointfinal} becomes
\begin{equation}\label{gaussianjointfinal}
\widehat{K}(x,y) = \exp\left[-\frac{a^2+b^2}{4 a^2 b^2}(x-y)^2\right] \int \frac{du}{\sqrt{(a^2+b^2)\pi}}\, \exp\left[-\frac{u^2}{a^2+b^2}\right]\, K(x+u,y+u)
\end{equation}
so that the disturbances
\begin{equation}
\eta(p)^2 =\frac{a^2+b^2}{2a^2b^2},\qquad \eta(x)^2 = \frac{a^2+b^2}{2}
\end{equation}
depend on the additional parameter, $a$.

One reasonable way to compare successive versus joint measurement, is to fix the disturbances to $x$ and $p$ (\emph{i.e.}, set the final states, \eqref{successivefinal} and \eqref{gaussianjointfinal}, equal), and then compare the measured uncertainties. This is possible when the detector resolutions for the successive measurements satisfy $\sigma_x\sigma_p\leq \tfrac{1}{4}$:
\begin{equation}
\begin{split}
a^2&= \frac{1}{4\sigma_p^2}\left(1+\sqrt{1-16\sigma_x^2\sigma_p^2}\right)\\
b^2&= \frac{1}{4\sigma_p^2}\left(1-\sqrt{1-16\sigma_x^2\sigma_p^2}\right)
\end{split}
\end{equation}
which yields the difference in measured uncertainties
\begin{subequations}
\begin{align}
(\Delta x)^2_{\text{joint}}-(\Delta x)^2_{\text{successive}}&= \frac{1}{16\sigma_p^2}\left(1-\sqrt{1-16\sigma_x^2\sigma_p^2}\right)^2\gt 0
\label{diffx}\\
(\Delta p)^2_{\text{joint}}-(\Delta p)^2_{\text{successive}}&=-\frac{1-4\sigma_x^2\sigma_p^2\left(1+\sqrt{1-16\sigma_x^2\sigma_p^2}\right)}{4\sigma_x^2\left(1-\sqrt{1-16\sigma_x^2\sigma_p^2}\right)}\lt 0\label{diffp}
\end{align}
\end{subequations}
The RHS of \eqref{diffp} ranges between $-\frac{1}{32\sigma_x^4\sigma_p^2}$ (for small $\sigma_x\sigma_p$) and $-\frac{3}{16\sigma_x^2}$ (for $\sigma_x\sigma_p=\tfrac{1}{4}$). So, in making a joint measurement, we trade a relatively modest deterioration in $(\Delta x)^2_{\text{measured}}$ for a potentially large improvement in $(\Delta p)^2_{\text{measured}}$.

Finally, note that Neumark's Theorem ensures that the POVM \eqref{AKpovm} for the joint measurement can be realized as a PVM in a larger Hilbert space, $\mathcal{H}\otimes\mathcal{H}'$ (in our case, we could choose $\mathcal{H}'=\mathcal{L}^2(\mathbb{R}^2)$). In this larger Hilbert space, the observables corresponding to this PVM \emph{commute}, as can be seen from \eqref{AKmoments}. Hence there exist, in this larger Hilbert space, states with arbitrary low values for the product of the corresponding quantum-mechanical uncertainties. For any $(\mathcal{H},A,B)$, it is always possible to find such an augmented Hilbert space, and pair of commuting observables thereon. Di Lorenzo \cite{DiLorenzo} would like to interpret this freedom as the ability to make arbitrary-precision measurements of the original (non-commuting) observables, $A,B$. But the states of the larger Hilbert space, in which the commuting observables have sharply-defined values, are highly-entangled (between the degrees of $\mathcal{H}$ and of $\mathcal{H}'$). The ability to prepare the ``system"+``measuring apparatus", in such a highly entangled initial state, belies the interpretation of this as a ``measurement."

\section{Relation with Ozawa}\label{OzawaRel}

\subsection{The disturbance $\eta(B)$ and Ozawa's uncertainty relation}\label{Ozawa}
Ozawa \cite{Ozawa,OzawaII} provides a more baroque definition of the disturbance, $\eta(B)$, than the one presented in \eqref{disturbance} above. Introduce an auxiliary Hilbert space, $\mathcal{H}'$ and a density matrix, $\chi$ on it. Let $U$ be a unitary operator on $\mathcal{H}\otimes\mathcal{H}'$, and let the above data ($\mathcal{H}',\chi,U$) be constrained by demanding that the density matrix $\widehat{\rho}$, after measuring $A$, be given by the partial trace,

\begin{displaymath}
\widehat{\rho}= Tr_{\mathcal{H}'}\left(U\, \rho\otimes \chi\, U^\dagger\right)
\end{displaymath}
Then Ozawa'{}s definition of $\eta(B)$ is

\begin{equation}
{\eta(B)}^2=Tr_{\mathcal{H}\otimes\mathcal{H}'}\left({(U^\dagger\, B\otimes\mathbb{1}\,U - B\otimes \mathbb{1})}^2\, \rho\otimes\chi\right)
\label{Ozawadef}\end{equation}
It seems rather uneconomical to introduce all this auxiliary data ($\mathcal{H}',\chi,U$) in order to define a quantity which should be intrinsic\footnote{It is important to emphasize that $\eta(B)$ \emph{should} be intrinsically-defined. Often, the auxiliary data ($\mathcal{H}',\chi,U$) is taken to be a crude model of the measurement process (with $\mathcal{H}'$ the Hilbert space of the measuring apparatus, $\chi$ its initial state, and $U$ the evolution operator for some time-dependent coupling of the system with the measuring apparatus). But any such model is a hopelessly-crude caricature of an \emph{actual} measuring apparatus; any results that depended on the details of the model would surely be inapplicable to any actual measuring apparatus. Instead, we adopt the philosophy that robustly-defined physical quantities depend only on the PVM/POVM, and not on any other auxiliary details.} to $(\mathcal{H},\rho, A,B)$. Indeed, there isn'{}t even a uniform prescription for making \emph{one} choice of ($\mathcal{H}',\chi,U$), given $(\mathcal{H},\rho, A,B)$.

If the number of distinct eigenvalues of $A$ is small enough, and if the Hilbert space is finite-dimensional, then there \emph{is} a uniform prescription that we can follow. Assume that $A$ has $N$ distinct eigenvalues, $\lambda_i$, with multiplicities $n_i$. Let

\begin{equation}
\mathcal{H}'=\mathcal{H},\quad \chi= \sum_{i=1}^N \frac{1}{n_i N} P^{(A)}_i
\label{Hprime}\end{equation}
which has the property that

\begin{displaymath}
Tr_{\mathcal{H}'}P^{(A)}_i\chi = \frac{1}{N},\quad\forall i
\end{displaymath}
Let

\begin{equation}
U = \exp\left[i\beta \sum_{i=1}^N P^{(A)}_i\otimes P^{(A)}_i\right]
\label{Udef}\end{equation}
We compute

\begin{displaymath}
\widehat{\rho} \equiv \sum_i P^{(A)}_i \rho P^{(A)}_i = Tr_{\mathcal{H}'}(U\,  \rho\otimes\chi\, U^\dagger)
\end{displaymath}
if and only if $\beta$ satisfies

\begin{displaymath}
2(1-\cos\beta) = N
\end{displaymath}
which has solutions for $N\leq 4$. Plugging \eqref{Hprime} and \eqref{Udef} into \eqref{Ozawadef}, we find

\begin{equation}
{\eta(B)}^2 = Tr\Bigl[\bigl((B-\widehat{B})^2 +\widehat{B^2}-\widehat{B}^2\bigr)\rho\Bigr]
\label{ourdef}\end{equation}
Even though we only computed \eqref{ourdef} for $N=2,3,4$, the result is $N$-independent and perfectly intrinsic. We take \eqref{ourdef} as a uniform definition, for all pairs of observables $A,B$, on any Hilbert space\footnote{
For any \emph{given} observable, $A$, we can always find, in a ad-hoc way, a triple, $(\mathcal{H}',\chi,U)$, which realizes the corresponding $\widehat{\rho}$, and then \emph{check} that \eqref{ourdef} is satisfied for any choice of $B$. When measuring $x$, as in \S\ref{successivexp}, we can take
\begin{itemize}%
\item $\mathcal{H}'=\mathcal{L}^2(\mathbb{R})$
\item $\chi$ given by the pure state with integral kernel,
$k(x',y')=f(x')\overline{f}(y')$, for some choice of acceptance function, $f(u)=f(-u)$.
\item $U$ implements a particular $SL(2,\mathbb{R})$ transformation. Let
\begin{displaymath}
\vec{x}=\left(\begin{smallmatrix}x\\ x'\end{smallmatrix}\right),\qquad R=\left(\begin{smallmatrix}1&0\\ -1&1
\end{smallmatrix}\right)
\end{displaymath}
Then $U:\psi(\vec{x})\mapsto \psi(R\vec{x})$.
\end{itemize}
Setting $x'=y'=x_0$, we reproduce \eqref{newK} and verify that \eqref{ourdef} holds for this case, too.}.

Having defined this peculiar observable, we might well ask, what does $\eta(B)$ \emph{mean}? There is one case where the answer is clear:

\begin{itemize}%
\item[] If $\widehat{B}=B$, then it follows that ${\langle B\rangle}_{\widehat{\rho}}={\langle B\rangle}_{\rho}$. (Note that the converse statement is \emph{not} true.) Hence, in this case,

\begin{displaymath}
\eta(B)^2 = {(\Delta B)}^2_{\widehat{\rho}}-{(\Delta B)}^2_{\rho}
\end{displaymath}
So $\eta(B)$ adds in quadrature to the quantum-mechanical uncertainty of $B$ in the initial state, ${(\Delta B)}_{\rho}$, to give the quantum-mechanical uncertainty in the final state, ${(\Delta B)}_{\widehat{\rho}}$.

\end{itemize}
More generally, $\eta(B)$ is unrelated to the change in the uncertainty of $B$ (which we saw, in \S\ref{twostate}, can even decrease). Nevertheless, $\eta(B)\geq0$. 

It is claimed in \cite{Ozawa} that $\eta(B)=0$ \emph{if and only if} the probability distribution for $B$ is the same in the state $\widehat{\rho}$ as it was in the initial state, $\rho$. That is \emph{obviously} false\footnote{The weaker statement:
\begin{itemize}\item[] \emph{$\eta(B)=0$ implies that the probability distribution for $B$ is unaltered.}\end{itemize}
may still be true. We have been unable to find either a counter-example or a proof of this weaker statement.}. In the 2-state system, we saw that, for $A=J_x$ and $B=\vec{b}\cdot\vec{J}$, $\eta(B)=\frac{1}{\sqrt{2}}\left(b_y^2+b_z^2\right)^{1/2}$ is completely independent of the state $\rho$. Take $\rho$ to be an eigenstate of $J_x$ (so that $\hat{\rho}=\rho$). The probability distribution for $B$ is unaffected by the measurement of $A$, even though we can \emph{clearly} have $\eta(B)>0$. 

Ozawa \cite{Ozawa} also proposes an uncertainty relation satisfied by $\eta(B)$. In our notation, it reads

\begin{equation}
\epsilon(A) \left(\eta(B) +{(\Delta B)}_\rho\right) +{(\Delta A)}_\rho \eta(B)\geq \frac{1}{2}\left|Tr([A,B]\rho)\right|
\label{bogus}\end{equation}
Here, $\epsilon(A)$ is the ``{}noise''{} in the measurement of $A$. For an observable, $A$, with a pure point spectrum, we can consider an ideal measuring device, with $\epsilon(A)=0$. However, for an observable, like $x$, with a continuous spectrum, we need to incorporate a finite detector resolution and $\epsilon(x)=\sigma_x$, the detector resolution we introduced previously. For $B=\mathbf{p}$, we computed in \eqref{best} that $\eta(p) \geq\frac{1}{2\sigma_x}$. Plugging this into \eqref{bogus} yields the uninteresting result

\begin{displaymath}
\sigma_x \left(\frac{1}{2\sigma_x} + {(\Delta p)}_\rho\right) + \frac{1}{2\sigma_x}{(\Delta x)}_\rho \geq \frac{1}{2}
\end{displaymath}
This ``{}lower bound''{} never comes close to being saturated. We computed a sharper lower bound in \eqref{xplowerbound}.

For the 2-state system, with $A=J_x$, $B=\vec{b}\cdot\vec{J}$ and $\rho=\frac{1}{2}(\mathbb{1}+\vec{a}\cdot\vec{\sigma})$, we computed

\begin{displaymath}
{(\Delta A)}_\rho = \frac{1}{2}{(1-a_x^2)}^{1/2},\quad \eta(B)=\frac{1}{\sqrt{2}}{(b_y^2+b_z^2)}^{1/2}
\end{displaymath}
and

\begin{displaymath}
\frac{1}{2} \left|Tr([A,B]\rho)\right| = \frac{1}{4}|b_y a_z-b_z a_y|
\end{displaymath}
Plugging these into \eqref{bogus}, we obtain

\begin{displaymath}
{(1-a_x^2)}^{1/2} {(b_y^2+b_z^2)}^{1/2}\geq \frac{1}{\sqrt{2}} |b_y a_z -b_z a_y|
\end{displaymath}
which, if we recall that $\vec{a}\cdot\vec{a}\leq 1$, is, indeed, satisfied, but is not the best bound one could obtain. In fact,

\begin{displaymath}
{(1-a_x^2)}^{1/2} {(b_y^2+b_z^2)}^{1/2}\geq |b_y a_z -b_z a_y|
\end{displaymath}

For the pure point spectrum case, we were able to consider the case $\epsilon(A)=0$. Now we turn to the case of a ``weak measurement" of $A$, which is one context where one naturally has $\epsilon(A)>0$.

\subsection{Weak Measurements, POVMs and $\epsilon(A)$}\label{Lund}

Lund and Wiseman \cite{Lund-Wiseman} propose a test of Ozawa'{}s ideas, using a ``{}weak measurement''{} of $A$. The notion of weak measurement introduced by Aharonov \emph{et al.}~\cite{PhysRevLett.60.1351} is slightly involved. However, in the case of a 2-state system discussed by Lund and Wiseman (and, more generally, for any observable $A$, such that $Spec(A)$ is a finite set), we can avail ourselves of a much simpler definition.

\begin{itemize}
\item[] A \emph{weak measurement of $A$} is a 1-parameter family of POVMs which interpolate between the PVM corresponding to $A$ and the uniform probability distribution on $Spec(A)$.
\end{itemize}

\noindent
More precisely, let $A$ have a pure point spectrum consisting of $N$ distinct eigenvalues, $\lambda_i$, and let $F_i(\theta)$ be a family of positive-semidefinite self-adjoint operators interpolating between

\begin{displaymath}
F_i(0) = P_i^{(A)} \quad \text{and}\quad F_i(\theta_{\text{max}})= \frac{1}{N} \mathbb{1}
\end{displaymath}
and satisfying

\begin{displaymath}
\sum_{i=1}^N F_i(\theta) = \mathbb{1}
\end{displaymath}
Concretely, let $f(\theta)$ be a monotonic function on $[0,\theta_{\text{max}}]$, satisfying

\begin{displaymath}
f(0)=1,\quad f(\theta_{\text{max}})= 0
\end{displaymath}
and let $F_i(\theta) = L_i^\dagger L_i$ with

\begin{equation}
L_i = \frac{1}{N}\Bigl(\sqrt{N-(N-1)f(\theta)}-\sqrt{f(\theta)}\Bigr)\mathbb{1} +\sqrt{f(\theta)} P^{(A)}_i
\label{Ldef}\end{equation}
Explicitly,

\begin{equation}
F_i(\theta) = \frac{1}{N}\bigl(1-g(\theta)\bigr)\mathbb{1}+ g(\theta) P^{(A)}_i
\label{FiDef}\end{equation}
where

\begin{equation}
g(\theta)= \left(1-\frac{2}{N}\right) f(\theta) +\frac{2}{N}\sqrt{N-(N-1)f(\theta)}\sqrt{f(\theta)}
\label{gdef}\end{equation}
von Neumann'{}s formula \eqref{povmcollapse} for the density matrix after measuring this POVM is

\begin{displaymath}
\widetilde{\rho} = \sum_i L_i \rho L_i^\dagger
\end{displaymath}
Using \eqref{Ldef}, we compute

\begin{displaymath}
\widetilde{\rho} = \bigl(1-f(\theta)\bigr) \rho + f(\theta) \widehat{\rho}
\end{displaymath}
which interpolates between $\widehat{\rho}$ (the change resulting from an ordinary measurement of $A$) and $\rho$ (no change)\footnote{To complete the prescription of a weak measurement \cite{Parrott}, one needs to define a set of \emph{contextual values} \cite{Dressel}, $\lambda_i(\theta)$, such that
\begin{displaymath}
A = \sum_i \lambda_i(\theta) F_i(\theta),\qquad \forall\, \theta\in[0,\theta_{\text{max}})
\end{displaymath}
With our parametrization,
\begin{displaymath}
\lambda_i(\theta) = \frac{1}{g(\theta)}\biggl(\lambda_i-\bigl(1-g(\theta)\bigr) \overline{\lambda}\biggr),\qquad \overline{\lambda}=\frac{1}{N}\sum_i\lambda_i
\end{displaymath}
}. Similarly, for any observable $O$, we can define

\begin{displaymath}
\widetilde{O} = \sum_i L_i O L_i^\dagger= (1-f(\theta)) O + f(\theta) \widehat{O}
\end{displaymath}
which interpolates between $\widehat{O}$ and $O$.

For the 2-state system, Lund and Wiseman made a specific choice,

\begin{displaymath}
f(\theta) = 1-\sin(2\theta), \qquad \theta_{\text{max}}= \pi/4
\end{displaymath}
which follows from an auxiliary triple $(\mathcal{H}',\chi,U)$ as in \S\ref{Ozawa}. Their choice (with a slight change in conventions ($\sigma_x\leftrightarrow\sigma_z$), since we want $A=J_x$ whereas they chose $A=2J_z$) was

\begin{displaymath}
\begin{gathered}
\mathcal{H}'=\mathcal{H}\\
\chi=\frac{1}{2}\bigl(\mathbb{1} +\cos(2\theta)\sigma_x+\sin(2\theta)\sigma_z\bigr)\\
U= \frac{1}{2}(1+\sigma_x)\otimes\mathbb{1}+\frac{1}{2}(1-\sigma_x)\otimes\sigma_z
\end{gathered}
\end{displaymath}
Of course, there'{}s nothing special about that choice; so long as we satisfy the constraint that

\begin{equation}
\widetilde{\rho} = Tr_{\mathcal{H}'} (U\, \rho\otimes\chi\, U^\dagger)
\label{tilderhoconstraint}\end{equation}
we can use any choice of $(\mathcal{H}',\chi,U)$. In fact, for $N\leq 4$, we can avail ourselves of the triple $(\mathcal{H}',\chi,U)$ introduced in \eqref{Hprime} and \eqref{Udef} by simply allowing $\beta$ in \eqref{Udef} to be $\theta$-dependent

\begin{displaymath}
\beta(\theta) = 2(\theta_{\text{max}}-\theta),\qquad \theta_{\text{max}} = \frac{2^{N-3}}{N}\pi
\end{displaymath}
which results in

\begin{displaymath}
f(\theta) = \frac{2}{N}\bigl(1-\cos\beta(\theta)\bigr)
\end{displaymath}
which agrees with Lund-Wiseman for $N=2$.

The disturbance in $B$, resulting from a weak measurement of $A$ can be defined similarly to \eqref{ourdef}

\begin{equation}
\eta_\theta(B)^2 = Tr\left[\left((B-\widetilde{B})^2+\widetilde{B^2}-\widetilde{B}^2\right)\rho\right]
\label{WeakDisturbance}\end{equation}
A simple calculation yields

\begin{displaymath}
\eta_\theta(B)^2 = f(\theta) \eta(B)^2
\end{displaymath}
with $\eta(B)^2$ given by \eqref{ourdef}. We also need to introduce $\epsilon(A)^2$, which represents the ``noise''{} inherent in a ``{}weak measurement''{} of $A$.

Lund and Wiseman define $\epsilon(A)^2$ in terms of a quadruple $(\mathcal{H}',\chi, U, M)$:

\begin{equation}
\epsilon(A)^2 = Tr_{\mathcal{H}\otimes\mathcal{H}'} \left[\left(U^\dagger\, \mathbb{1}\otimes M\, U - A\otimes \mathbb{1}\right)^2 \rho\otimes\chi\right]
\label{LWdef}\end{equation}
where $\mathcal{H}',\chi, U$ are as before and $M$ is \emph{some} self-adjoint operator on $\mathcal{H}'$. Unfortunately, aside from the obvious requirement that \eqref{tilderhoconstraint} be satisfied, this ``{}definition''{} is a little \ldots{} \emph{under-specifed}. Instead, we directly implement the notion that $\epsilon(A)^2$ represents the mean-squared deviation of the weakly-measured $A$ from its true value by writing

\begin{equation}
\epsilon(A)^2 = \sum_{i,j} (\lambda_i-\lambda_j)^2 Tr\left(F_i(\theta) P^{(A)}_j \rho\right)
\label{ourepsilon}\end{equation}
Using \eqref{FiDef}, this reduces to

\begin{displaymath}
\epsilon(A)^2 = \frac{1}{N} \bigl(1-g(\theta)\bigr)\sum_{i,j} (\lambda_i-\lambda_j)^2 Tr\left(P^{(A)}_j \rho\right)
\end{displaymath}
which agrees with their explicit realization of \eqref{LWdef} for the 2-state system (the only case we will actually need).

Putting these together, we obtain the Ozawa/Lund-Wiseman uncertainty relation for a \emph{weak} measurement of $A=J_x$, followed by a measurement of $B=\vec{b}\cdot \vec{J}$, starting with the initial state $\rho = \frac{1}{2} \left(\mathbb{1}+\vec{a}\cdot\vec{\sigma}\right)$, is:

\begin{equation}
\begin{split}
2\sin(\theta)\bigl(\cos(\theta)-\sin(\theta)\bigr)\left(b_y^2+b_z^2\right)^{1/2}
+\sqrt{2}\sin(\theta)\left(b^2-(\vec{a}\cdot\vec{b})^2\right)^{1/2} &+\\
 + \left(1-a_x^2\right)^{1/2} \left(b_y^2+b_z^2\right)^{1/2}
&\geq \frac{1}{\sqrt{2}} \left|b_y a_z - b_z a_y\right|
\end{split}
\label{OLW}\end{equation}
In \S\ref{Ozawa}, we already saw that this inequality was ineffective (never close to being saturated) at $\theta=0$. Here, we see that the situation is even worse at nonzero $\theta$, since the ``{}new''{} terms on the LHS are non-negative for $0\lt\theta\lt \pi/4$.

Rozema \emph{et al}.~\cite{Rozema:2012sg}, in a clever experiment, measure the quantities appearing on the LHS and RHS of \eqref{OLW} separately. Indeed, their results for the LHS nowhere approach the values for the RHS.

Finally, let us briefly return to the subject of \S\ref{successivexp}. Measuring $x$, with finite detector resolution $\sigma_x$, is naturally associated to the POVM
\begin{displaymath}
S\mapsto \mathcal{F}_S = \int_S d x_0\,  F_{x_0}
\end{displaymath}
The analogue of \eqref{ourepsilon} is 
\begin{equation}\label{epsilonx}
\begin{split}
\epsilon(x)^2 &=  \int d x_0\,  Tr(F_{x_0}\, (\mathbf{x}-x_0\mathbb{1})^2\rho)\\
&= Tr(\mathbf{x}^2\rho)+ \int d x_0\left[ x_0^2\, Tr(F_{x_0}\rho) -2 x_0\, Tr(F_{x_0}\mathbf{x}\rho) \right]
\end{split}
\end{equation}
and a simple computation yields
\begin{displaymath}
\epsilon(x)^2 = \sigma_x^2
\end{displaymath}
as expected.
 
\section*{Acknowledgements}\label{acknowledgements}
The research of the authors is based upon work supported by the National Science Foundation under Grant No. PHY-0969020. We would like to thank Antonio~Di Lorenzo, Dan Freed, Matt Headrick and Steve Weinberg for useful discussions. We would particularly like to thank the latter for encouraging us to write up these observations.

\vfill\eject
\bibliographystyle{utphys}
\bibliography{uncertainty}

\end{document}